# Terahertz Plasmonic Detector Controlled by Phase Asymmetry

I. V. Gorbenko[1], V. Yu. Kachorovskii[1*] and Michael Shur[2,3*]

[1]*A. F. Ioffe Physico-Technical Institute, St. Petersburg 194021, Russia*
[2] *Rensselaer Polytechnic Institute, Troy, NY 12180, USA*
[3] *Electronics of the Future, Inc., Vienna, VA 22181, USA*
**\*kachor.valentin@gmail.com*

**Abstract:** We demonstrate that phase-difference between terahertz signals on the source and drain of a field effect transistor (a TeraFET) induces a plasmon-assisted dc current, which is dramatically enhanced in vicinity of plasmonic resonances. We describe a TeraFET operation with identical amplitudes of radiation on source and drain antennas but with a phase-shift-induced asymmetry. In this regime, the TeraFET operates as a *tunable resonant polarization-sensitive* plasmonic spectrometer operating in the sub-terahertz and terahertz range of frequencies. We also propose an effective scheme of a phase-sensitive homodyne detector operating in a phase-asymmetry mode, which allows for a dramatic enhancement of the response. These regimes can be implemented in different materials systems including silicon. The p-diamond TeraFETs could support operation in the 200 to 600 GHz atmospheric windows.

**1. Introduction**

The TeraFETs [1, 2] – the devices transforming terahertz (THz) radiation into dc current by using the excitation of overdamped or resonant plasma oscillations implemented in Si [3, 4], GaAs [5, 6], GaN [7], graphene [8], and other materials systems are now being commercialized as tunable and fast detectors of sub-THz and THz radiation. The underlying physical mechanism proposed in Refs [1,2] is based on the rectification of plasmonic oscillations enabled by the device nonlinearity and the asymmetry of the setup. The latter determines the direction of the dc current response (or sign of the induced dc voltage). One of the possible ways to introduce the asymmetry is to impose different boundary conditions on the source and drain of the device. In particular, when the source-gate input is excited by a signal with amplitude $U_a$, while the drain current is fixed the generated open circuit dc voltage is

$$V \propto U_a^2.$$

Such drain boundary condition implicitly implies that the external circuit is connected to the drain via an antenna or a contact pad with an infinite inductive impedance. Special "plasmonic stub" structures [13] might help implement such a boundary condition. Another option is to use identical antennas but apply signals with different amplitudes $U_a$ and $U_b$ at the source and drain, respectively. For such setup, the induced dc current is proportional to the difference of the squared voltages at the source and drain (assuming that these signals have the same phase) [9]

$$V \propto U_a^2 - U_b^2$$

Importantly, a strong asymmetry can be also induced by the phase shift between the THz signals at the source and drain. For equal amplitudes of signals on the source and drain $U_a = U_b$ nonzero phase shift $\theta$ between signals induces dc current [10-12]

$$V \propto U_a^2 \sin\theta$$

In [10-12], the circularly-polarized radiation was used to induce such a phase shift. Remarkably, this shift depends very weakly on $L/\lambda$ (here $L$ is the channel length and $\lambda$ is the THz radiation wavelength) and remains finite even in the limit of homogeneous field. In this limit, the phase shift is simply given by geometrical angle $\pm\theta$ (see Fig.1 a) with the sign being determined by the clockwise or counterclockwise helicity.

Another possibility is to use the identical antennas ($\theta = 0$ in Fig 1 a) but change the angle between incoming radiation and the plane of the device. In this case, one can use a linearly polarized radiation.

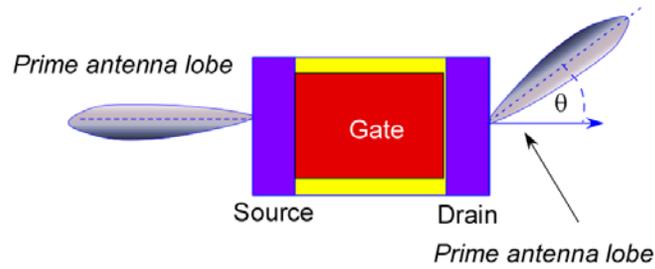

(a)

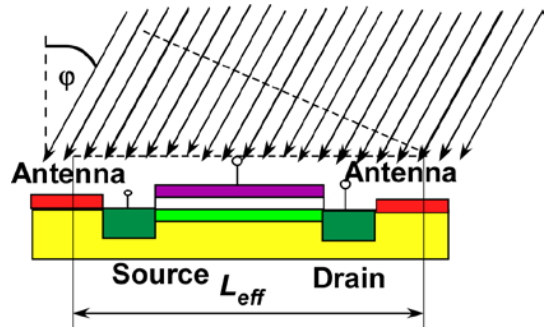

(b)

Fig. 1. TeraFET Spectrometer principle of operation: (a) phase shift induced by asymmetric antennas and circularly polarized radiation (b) nonzero incident angle of incoming radiation

Based on this physical idea, we show below that a single transistor can be used as spectrometer of the THz or sub-THz radiation.-The presented results reveal the new physics of the TeraFET spectrometer and provide the design, characterization and parameter extraction tools for the new generation of the phase sensitive interferometer/spectrometers.

We will also discuss the homodyne detection scheme [14] enabled by the phase asymmetry. In this scheme, a strong local oscillator signal with the amplitude $U_{\text{local}}$ symmetrically excites the source and drain, while the incoming radiation with the amplitude $U_{\text{signal}} \ll U_{\text{local}}$ induces the phase-shifted (by the phase $\theta$) signals on opposite sides of the channel. As we will show, the dc response in this case reads

$$V \approx U_{\text{local}} U_{\text{signal}} \left[ A(1 - \cos\theta) + B\sin\theta \right] \tag{1}$$

where coefficients A and B have qualitatively different frequency dependencies.

## 2. TeraFET Spectrometer Principle of Operation

Fig. 1 showing the TeraFET spectrometer structure illustrates its principle of operation. A THz radiation impinging on the FET couples to the two antennas at the opposite sides of the channel. The asymmetry is caused by the phase shift $\theta$ between these antennas. This shift depends on the polarization of the radiation and the geometry of the setup. For a circular polarization, $\theta$ is nonzero provided that antennas configuration is asymmetric with respect to direction from the source to drain. For a linear polarization, the finite phase shift $\theta \propto \sin\varphi$ appears for nonzero incidence angle $\varphi$ (coefficient in this equation depends on the geometry detail). Importantly, this phase shift enters the boundary conditions for the electron fluid in the transistor channel

$$\begin{aligned} U(0) &= U_g + U_a \cos(\omega t) \\ U(L) &= U_g + U_a \cos(\omega t + \theta) \end{aligned} \tag{2}$$

Here $U(0)$ and $U(L)$ are the voltages at the source and drain of the channel, respectively, $U_a$ is the THz induced voltage amplitude, $U_g$ is the gate-to-channel voltage swing (counted form the threshold voltage) and $\omega$ is the frequency of the impinging THz radiation. The electron fluid is described by the standard hydrodynamic equations

$$\frac{\partial v}{\partial t} + v\frac{\partial v}{\partial x} + \gamma v = -\frac{e}{m}\frac{\partial U}{\partial x} \tag{3}$$

$$\frac{\partial U}{\partial t} + \frac{\partial (Uv)}{\partial x} = 0 \tag{4}$$

Here $v$ is the velocity of the fluid, $U$ is the local value of the gate-to-channel voltage swing. related to the electron concentration in the channel:

$$n_s = CU/e, \tag{5}$$

$\gamma$ is the inverse momentum relaxation time, The solution of Eqs. (3-5) with the boundary conditions given by Eq. (2) is obtained using the same approach as in [2,12]

$$V = \frac{\beta \omega U_a^2 \sin\theta}{4U_g |\sin(kL)|^2 \sqrt{\omega^2 + \gamma^2}}, \qquad (6)$$

Here $\beta = 8\sinh\left(\frac{\Gamma L}{s}\right)\sin\left(\frac{\Omega L}{s}\right)$, $k = (\Omega + i\Gamma)/s$, and

$$\Omega = \sqrt{\frac{\sqrt{\omega^4 + \omega^2\gamma^2}}{2} + \frac{\omega^2}{2}}; \quad \Gamma = \sqrt{\frac{\sqrt{\omega^4 + \omega^2\gamma^2}}{2} - \frac{\omega^2}{2}} \qquad (7)$$

Here $k$ is the wave vector, $\omega$ is frequency, $\Omega$ is the plasma frequency, $\Gamma$ is the effective damping rate, $s = \sqrt{eU_g/m}$ is the plasma wave velocity, $e$ is the electron charge, and $m$ is the effective mass. In the resonant regime $\omega \gg \gamma$, $s/L \gg \gamma$ these equations simplify

$$\Omega \approx \omega, \; \Gamma \approx \frac{\gamma}{2}, \; \Omega \gg \Gamma, \qquad (8)$$

and one finds a sharply-peaked response at resonant frequencies $\omega_N = \pi N s/L$:

$$V \approx \frac{4\delta\omega\gamma(-1)^N}{4U_g(\delta\omega^2 + \gamma^2/4)} U_a^2 \sin\theta, \qquad (9)$$

where $\delta\omega = \omega - \omega_N \ll \omega_N$. We notice an asymmetric form of the resonances, $V(\delta\omega) = -V(-\delta\omega)$, in contrast to the symmetric resonances obtained under the open drain boundary conditions [2].

Both in the resonant and non-resonant case, Eq. (6) shows a periodic variation with frequency. In particular, the response exactly turns to zero for $\Omega = \Omega_N = \pi s N/L$ due to factor $\sin(\Omega L/s)$ in the coefficient $\beta$. These results enable the application as a spectrometer. This spectrometer operates as follows. At each frequency, the gate-to-source voltage will be adjusted till the response is zero at every incidence angle. This yields the values of the frequency satisfying the following condition

$$\sqrt{\frac{\sqrt{\omega_N^4 + \omega_N^2\gamma^2}}{2} + \frac{\omega_N^2}{2}} = \frac{\pi s}{L} N \qquad (10)$$

Here $N=1, 2, 3\ldots$. For a monochromatic signal with a frequency $\omega$, one can tune $\omega_N$ to the value $\omega$ by changing $s$ by the gate voltage. This allows to measure $\omega$. For a more general case of the radiation with a spectrum broadened around $\bar{\omega}$ within a certain interval $\Delta\omega$ with the wave amplitude given by $U_a(\omega)$, the dc response can have found by replacing in Eq. 6 $U_a \to U_a(\omega)$ and integrating over $\omega$. Tuning the resonant frequency $\omega_N$ to cover interval $\bar{\omega} \pm \Delta\omega$ by the gate voltage; one can extract the spectral density $|U_a(\omega)|^2$.

## 3. Homodyne detector operation scheme

Let us now assume that. in addition to the phase-shifted signal of the incoming radiation, we apply a strong fully symmetric signal of a local oscillator. This situation can be modeled by the following boundary conditions (see Fig. 2)

$$U(0) = U_{local} \cos \omega t + U_{signal} \cos \omega t$$
$$U(L) = U_{local} \cos \omega t + U_{signal} \cos(\omega t + \theta) \tag{11}$$

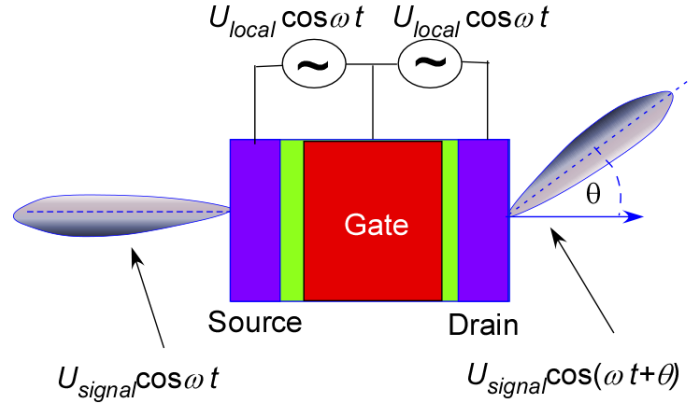

Fig. 2. Homodyne detector operation scheme

These equations can be rewritten in a form used in Ref. [12]

$$U(0) = U_a \cos \omega t,$$
$$U(L) = U_b \cos(\omega t + \tilde{\theta}), \tag{12}$$

where parameters $U_a, U_b$ and $\tilde{\theta}$ can be found from the following equations $U_a = U_{local} + U_{signal}$, $U_b e^{i\tilde{\theta}} = U_{local} + U_{signal} e^{i\theta}$, From these equations we find that in the presence of the local oscillator, the phase asymmetry of the incoming signal induces effective asymmetry of the amplitudes. For $U_{local} \gg U_{signal}$, we find $U_a^2 - U_b^2 \approx 2U_{local}U_{signal}(1 - \cos\theta)$. We also find $U_a U_b \sin\tilde{\theta} \approx U_{local}U_{signal}\sin\theta$. The calculations analogous to [2,12] yield

$$A = \frac{\omega}{2U_g |\sin(kL)|^2 \sqrt{\omega^2 + \gamma^2}} \left[ \left(1 + \frac{\gamma\Omega}{\Gamma\omega}\right)\sinh^2\left(\frac{\Gamma L}{s}\right) - \left(1 - \frac{\gamma\Gamma}{\omega\Omega}\right)\sin^2\left(\frac{\Omega L}{s}\right) \right],$$
$$B = \frac{2\omega}{U_g |\sin(kL)|^2 \sqrt{\omega^2 + \gamma^2}} \sinh\left(\frac{\Gamma L}{s}\right)\sin\left(\frac{\Omega L}{s}\right). \tag{13}$$

In the resonant regime $\omega \gg \gamma$, $s/L \gg \gamma$, we find from Eqs. 1 and 13

$$V \approx \frac{U_{local}U_{signal}}{2U_g} \frac{(3\gamma^2/4 - \delta\omega^2)(1-\cos\theta) + 2(-1)^N \gamma\delta\omega\sin\theta}{(\delta\omega^2 + \gamma^2/4)}. \quad (14)$$

As seen, there are two contributions to the response having qualitatively different frequency dependencies in vicinity of plasmonic resonances: the symmetric term, which is proportional to $(1-\cos\theta)$ and the asymmetric one, which is proportional to $\sin\theta$. For $\delta\omega \to \infty$ response remains finite. Extracting value of response at $\delta\omega = \infty$, we get

$$V(\delta\omega) - V(\infty) = \frac{U_{local}U_{signal}}{2U_g} F\left(\frac{\delta\omega}{\gamma}\right), \quad (15)$$

where function

$$F(x) = \frac{1 - \cos\theta + 2(-1)^N x \sin\theta}{x^2 + 1/4} \quad (16)$$

is plotted in Fig. 3.

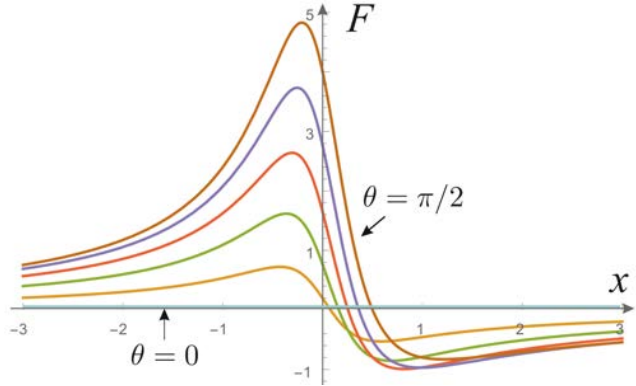

**Figure 3** Resonant dependence of dimensionless response on the radiation frequency [Eq. (16)] for fundamental plasmonic frequency ($N = 1$) at different $\theta$ ($\theta/\pi = 0.1, 0.2, 0.3, 0.4, 0.5$) increasing from bottom to the top at negative $x$.

## 4. Calculation of response for various materials

Next, we present the response calculations for different materials. Figures 4-7 show the calculation results for p-diamond FETs, Si NMOS, AlGaN/GaN and InGaAs/InP HEMTs, respectively, for the channel lengths of 25 nm, 65 nm, and 130 nm (the 250 nm results are also shown for p-diamond). Table 1 lists the parameters used in the calculation.

TABLE I.    MATERIALS PARAMETERS

| Material | Parameters | | |
|---|---|---|---|
| | Effective mass, $m_r$ | Mobility ($m^2/Vs$) | Scattering Frequency, $\gamma$ ($10^{12}/s$) |
| p-diamond | 0.74 | 0.53 | 0.45 |
| n-Si | 0.19 | 0.10 | 9.3 |
| n-GaN | 0.24 | 0.15 | 4.9 |
| n-InGaAs | 0.041 | 0.8 | 5.4 |

As seen from Fig. 4 a, the p-diamond FETs should enable room temperature spectroscopy in the sub-THz range. Other materials systems also offer unique capabilities for THz spectroscopy.

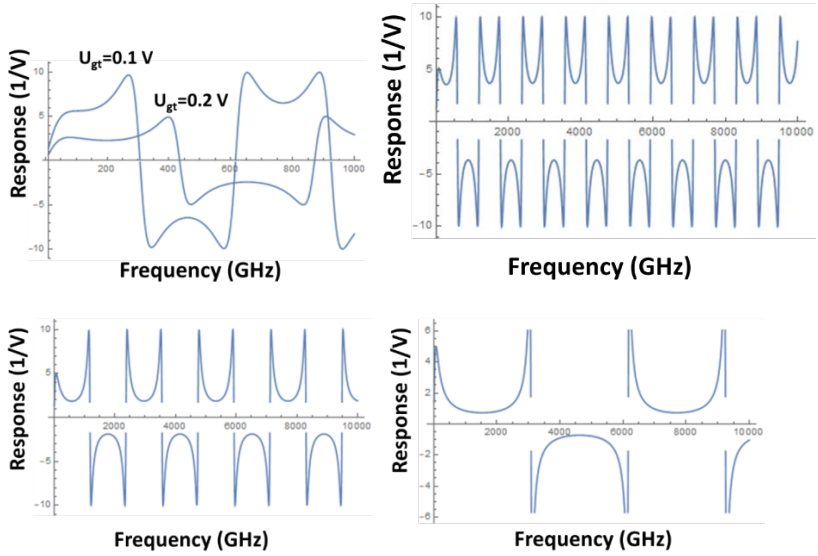

Fig. 4. Diamond TeraFET response normalized to $U_a^2$ for 250 nm (a), 130 nm (b), 65 nm (c) and 25nm (d) channel lengths

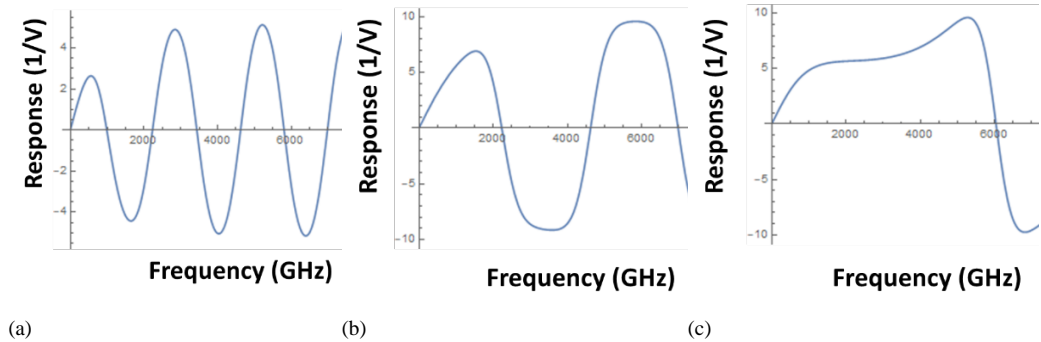

(a)                     (b)                     (c)

Fig. 5. Silicon TeraFET response normalized to $U_a^2$ for 130 nm (a), 65 nm (b) and 25nm (c) channel lengths.

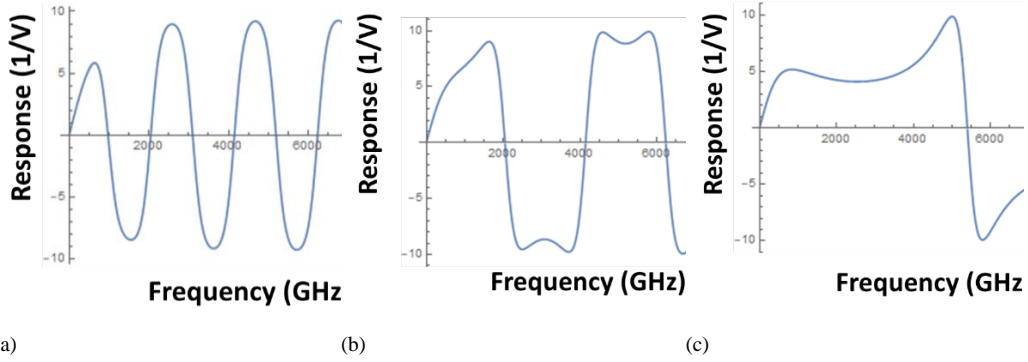

(a)                    (b)                    (c)

Fig. 6. AlGaN/GaN TeraFET response normalized to $U_a^2$ for 130 nm (a), 65 nm (b) and 25nm (c) channel lengths.

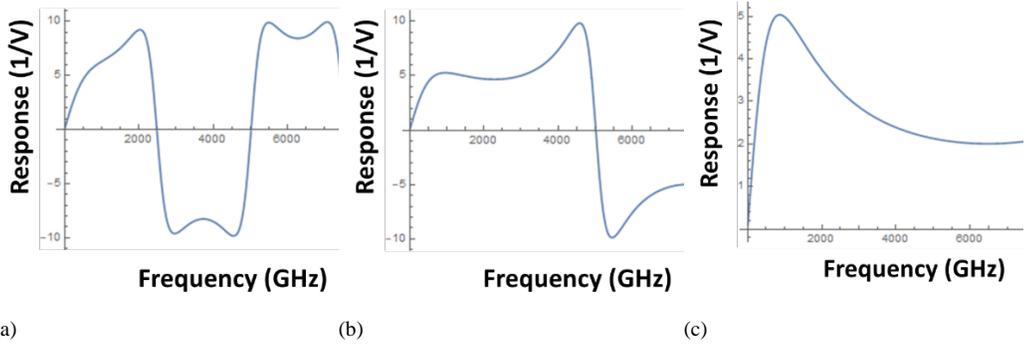

(a)                    (b)                    (c)

Fig. 7. AlGaAs/GaAs TeraFET response normalized to $U_a^2$ for 130 nm (a), 65 nm (b) and 25nm (c) channel lengths.

As seen, the responses for all these materials strongly depend on frequency even at for the parameters corresponding to room temperature. Compared to a conventional TeraFET detector, the approach based on the phase being responsible for the asymmetry is much easier to control because it eliminates issues related to the frequency dependent input, output, and load impedances, since it could be realized using identical antennas at the source and drain of the TeraFET. Even silicon TeraFETs seem to be a reasonable option, which allows to interface the TeraFET with inexpensive standard VLSI processing and data acquisition hardware. The results for p-diamond [diamond] confirm a big potential of this material for applications in the 240-320 GHz and 500 to 600 GHz atmospheric windows (the potential frequency ranges for beyond 5G).

For resonant homodyne regime of operation response is given by Eq. (15) where $\gamma$ for different materials are given in Table I.

## 5. Conclusions

To conclude, we developed a theory of phase-asymmetry-induced dc photoresponse in the FET subjected to THz radiation. We found that response shows sharp resonant peaks in vicinity of plasmonic resonances. The peaks have asymmetric shape as a function of frequency. We also discussed the homodyne operation, where response is drastically increased by using a strong local oscillator signal. In this case, the response contains two terms with symmetric and asymmetric frequency dependencies. These results can be used for creation compact, tunable spectrometers of THz radiation. We present detailed calculations of the response for different materials with different mobilities and find sufficiently relaxed conditions for realization of the resonant response.

**5. Acknowledgment**

The work of M. S. S. was supported by the U.S. Army Research Laboratory through the Collaborative Research Alliance for Multi-Scale Modeling of Electronic Materials and by the Office of the Naval Research (Project Monitor Dr. Paul Maki). The work of I.V.G. and V. Yu. K. was supported by the Foundation for the advancement of theoretical physics "BASIS" and by the RFBR grant 17-02-00217.

**References**

1. M. Dyakonov and M. S. Shur, Phys. Rev. Lett. 71, 2465 (1993).
2. M. Dyakonov and M.S. Shur, IEEE Transaction on Electron Devices 43, 380 (1996).
3. W. Knap, F. Teppe, Y. Meziani, N. Dyakonova, J. Lusakowski, F. Boeuf, T. Skotnicki, D. Maude, S. Rumyantsev, M. Shur, Appl. Phys. Lett. 85, 675 (2004).
4. R. Tauk, F. Teppe, S. Boubanga, D. Coquillat, W. Knap, Y. M. Meziani, C. Gallon, F. Boeuf, T. Skotnicki, C. Fenouillet-Beranger, D. K. Maude, S. Rumyantsev and M. S. Shur, Appl. Phys. Lett. 89, 253511 (2006).
5. W. Knap, V. Kachorovskii, Y. Deng, S. Rumyantsev, J.Q. Lu, R. Gaska, M.S. Shur, G. Simin, X. Hu and M. Asif Khan, C.A. Saylor, L.C. Brunel, J. Appl. Phys. 91, 9346 (2002).
6. F. Teppe, M. Orlov, A. El Fatimy, A. Tiberj, W. Knap, J. Torres, V. Gavrilenko, A. Shchepetov, Y. Roelens, and S. Bollaert, Appl. Phys. Lett. 89, 222109 (2006).
7. A. El Fatimy, S. Boubanga Tombet, F. Teppe, W. Knap, D.B. Veksler, S. Rumyantsev, M.S. Shur, N. Pala, R. Gaska, Q. Fareed, X. Hu, D. Seliuta, G. Valusis, C. Gaquiere, D. Theron, and A. Cappy, Electronics Letters, 42, 1342 (2006).
8. T. Otsuji, V. Popov, and V. Ryzhii, Journal of Physics, D: Applied Physics, 47, 094006 (2014). 315-320
9. D. Veksler, A. Muraviev, V.Yu. Kachorovskii, T. Elkhatib, K. Salama, X.-C Zhang, M. Shur, Solid-State Electronics. 53. 571-573. (2009)
10. C. Drexler, N. Dyakonova, P. Olbrich, J. Karch, M. Schafberger, K. Karpierz, Yu. Mityagin, M. B. Lifshits, F. Teppe, O. Klimenko, Y. M. Meziani, W. Knap, and S. D. Ganichev, J. Appl. Phys. 2012, 111, 124504.
11. K. S. Romanov and M. I. Dyakonov, Applied Phys. Lett.2013, 102, 153502.
12. I.V. Gorbenko, V.Yu. Kachorovskii, and M.S. Shur, PSS, Rapid Research Letters, in press, DOI: 10.1002/pssr.201800464 (2018)
13. G. R. Aizin, J. Mikalopas, M. Shur, https://arxiv.org/abs/1806.00682, 2018 IEEE ED, accepted, 10.1109/TED.2018.2854869
14. S. Rumyantsev, and V, Kachorovskii, and M. Shur, Homodyne Phase Sensitive Terahertz Spectrometer, Appl. Phys. Lett., 111, 121105 (2017), doi: 10.1063/1.5004132S